\documentclass[12pt]{article}
 \usepackage{verbatim}
 \usepackage{geometry}
 \usepackage{amssymb}
 \usepackage{units}
 \usepackage{setspace}
 \usepackage{dsfont}
  \usepackage[numbers]{natbib}
   \usepackage{color}
   \usepackage{amsmath}
   \usepackage{graphicx}

\title{Dose Finding Studies for Therapies with Late-Onset Safety and Efficacy Outcomes}

\author{Helen Barnett$^1$, Oliver Boix$^2$, Dimitris Kontos$^3$, Thomas Jaki$^{1,4}$\\$^1$ MRC Biostatistics Unit, University of Cambridge\\
$^2$ Bayer AG\\
$^3$ ClinBAY\\
$^4$ Department of Mathematics and Statistics, Lancaster University}
\begin{document}
\maketitle
\begin{abstract}
In Phase I/II dose-finding trials, the objective is to find the Optimal Biological Dose (OBD), a dose that is both safe and efficacious that maximises some optimality criterion based on safety and efficacy. This is further complicated when the investigated treatment consists of multiple cycles of therapy, and both toxicity and efficacy outcomes may occur at any point throughout the follow up of multiple cycles. In this work we present the Joint TITE-CRM, a model-based design for late onset toxicities and efficacies based on the well-known TITE-CRM. It is found to be superior to both currently available alternative designs that account for late onset bivariate outcomes; a model-assisted method and a bivariate survival design, as well as being both intuitive and computationally feasible.
\end{abstract}

\section{Introduction} \label{sec:intro}

In traditional drug development, safety and efficacy of potential new drugs have been evaluated separately. A phase I trial first finds the maximum tolerated dose, the dose associated with some predetermined probability of observing a Dose-Limiting Toxicity (DLT). This dose is then carried forward to phase II, where efficacy is evaluated, with limited borrowing of information between the two evaluations of safety and efficacy. An alternative option is a seamless Phase I/II trial where safety and efficacy are evaluated simultaneously, with the aim to find the Optimum Biological Dose (OBD). The main advantage of collecting information on both is the increase in chances of finding a dose that is both safe and efficacious, by allowing for more sharing of information. \\

There are multiple methods that have been proposed to design such a trial that range in complexity, for example the relatively simple model-assisted toxicity and efficacy interval design of Lin and Yin (STEIN, \citep{Lin2017}), or more complex model-based designs that can jointly model the efficacy-dose and toxicity-dose relationship such as the utility contour of Thall and Cook \citep{Thall2004}, or the approach based on toxicity and efficacy odds ratios by Lin et al. \citep{Yin2006}, or the bivariate CRM for competing outcomes of Braun \citep{Braun2002} amongst others. These methods differ in both their approaches to inference of the bivariate (or trinary) outcomes, and the decision criteria based on this inference. A standard trial setting assumes that as each cohort of patients enters the trial, responses from all previous cohorts are available to inform the decision of the next dose assignment. The next cohort is assigned the dose in order to collect more information at the current estimate of the OBD.\\

However, it is not always the case that the complete outcomes of all previous cohorts are available before the next cohort is assigned. Particularly in Oncology, where there are multiple cycles of treatment, both efficacy and safety outcomes may have a delayed onset. There are a limited number of phase I designs that account for late onset toxicities, but an even more limited number of designs that account for late onset outcomes in both safety and efficacy. The main contributions to dose-finding trial designs incorporating late onset toxicities include the interval censored approach of Sinclair and Whitehead (ISCDP, \citep{Sinclair2014}), the approach of including a cycle effect in a proportional odds mixed effects model by Dousseau et al. (POMM, \citep{Doussau2013}) and the Time-to-Event version of the Continual Reassessment Method by Cheung and Chappell (TITE-CRM, \citep{YingKuenCheung2000}). When including late-onset efficacy as well, there is a limited number of approaches, for example the method jointly modelling the time-to-event of efficacy and toxicity of Yuan and Yin ($A_T/A_E$, \citep{Yuan2009}), or the model assisted approach of Liu and Johnson \citep{Liu2016}.\\

The setting of late onset toxicities and efficacies is challenging, hence the very few methods that exist for such a trial, yet extremely relevant when multiple cycles of treatment are given. Using one of the many available methods for binary outcomes will give a much longer trial duration, as all cycles of treatment must be completed before the next dose assignment.\\

In this paper we investigate a joint time to event CRM for the implementation of a phase I/II trial design with delayed onset outcomes for both safety and efficacy, in comparison to the existing methods. In Section~\ref{sec:methods} we introduce the methods and setting, presenting results of their application in Section \ref{sec:results} before concluding with a discussion in Section~\ref{sec:discussion}.\\
\section{Methods} \label{sec:methods}
We consider the setting where there are $J$ dose levels with values $d_1, \ldots , d_j, \ldots , d_J$. Patients enter the trial in cohorts. Each cohort is assigned to a dose, and followed up for $\tau$ cycles of treatment. A DLT may occur at any time during the follow up period, in which case the patient leaves the trial. The patient may also observe an efficacy response at any time during their follow up, which is censored if the patient observes a DLT.\\

The choice of dose to assign to the next cohort, and the final dose recommendation, are chosen based on the design. Here we outline the proposed Joint TITE-CRM design, the $AT/AE$ design of Yuan \& Yin \citep{Yuan2009} and the model-assisted design of Liu \& Johnson \citep{Liu2016}. For comparability, in each case, the trial proceeds in the following way\\
\begin{enumerate}
\item The first cohort of patients is assigned to the lowest dose.
\item After one cycle of follow up, if no DLT is observed, escalate to next highest dose and continue after each cycle until a DLT is observed or the highest dose is reached.
\item The relevant models are fitted to the currently observed responses (which will be from one cycle of follow up for the last cohort, two cycles for the last but one cohort etc.) and posterior distributions are updated. \label{item:pp}
\item A set of `admissible' doses are calculated based on the models fitted.
\item The best dose of the `admissible' set according to some criterion of the design is chosen to assign to the next cohort (subject to certain prespecified rules discussed in Section~\ref{sec:rules}). \label{item:crit}
\item After one cycle of follow up, return to step \ref{item:pp}. The trial is stopped when one of the pre-specified stopping rules is triggered. 
\end{enumerate}

It is worthwhile to note that the form of the prior and posterior in step \ref{item:pp} and criterion in step \ref{item:crit} are the only aspects unique to each design.

\subsection{Joint Model TITE-CRM} \label{sec:j_titecrm}
Based on the TITE-CRM \citep{YingKuenCheung2000} and similar to the bCRM of Braun \cite{Braun2002}, we modify the procedure to include both safety and efficacy, whilst keeping the structure of the TITE-CRM.\\

We use a two-parameter logistic model for both efficacy and toxicity outcomes:
\[
F(d,\mathbf{\beta_E})= \frac{exp(\beta_{E,0} + \beta_{E,1} d)}{1+exp(\beta_{E,0}  + \beta_{E,1} d)},
\]
\[
F(d,\mathbf{\beta_T})= \frac{exp(\beta_{T,0} + \beta_{T,1} d)}{1+exp(\beta_{T,0} + \beta_{T,1} d)},
\]

We then must give weights to both efficacy and toxicity observations based on their follow ups, and a weighted dose response model is used for each:
\[ 
{G(d,w{(E)},\beta_E) = w^{(E)} F(d,\beta_E)},
\]
\[ 
{G(d,w{(T)},\beta_T) = w^{(T)} F(d,\beta_T)},
\]
where {$0 \leq w^{(E)}, w^{(T)} \leq 1$} are functions of time-to-event of a patient response. We use $w^{(T)}_{i,n}=u_{i,n}/\tau$, where $u_{i,n}$ is the current number of cycles patient $i$ has been observed for, unless a DLT is observed in which case $w^{(T)}_{i,n}=1$. In a similar fashion, we use $w^{(E)}_{i,n}=u_{i,n}/\tau$, where $u_{i,n}$ is the current number of cycles patient $i$ has been observed for, unless a DLT is observed before an efficacy outcome can be observed in which case $w^{(E)}_{i,n}=(\mbox{DLT time - entry time})/\tau$, and $w^{(E)}_{i,n}=1$ if an efficacy outcome is observed.\\

We consider all binary combinations of efficacy and toxicity outcomes (as opposed to trinary sometimes used in such applications), related by a Gumbel Model \citep{Thall2004}

\begin{align*}
\pi_{a,b}= &(G(d,w{(E)},\beta_E) )^a (1-G(d,w{(E)},\beta_E) )^{1-a} (G(d,w{(T)},\beta_T) )^b (1-G(d,w{(T)},\beta_T) )^{1-b} + \\ &(-1)^{a+b} (G(d,w{(E)},\beta_E) ) (1-G(d,w{(E)},\beta_E) ) G(d,w{(T)},\beta_T)  (1-G(d,w{(T)},\beta_T) ) \left( \frac{e^{\psi}-1}{e^{\psi}+1} \right),
\end{align*}

where $\pi_{a,b}$ is the probability of observing the binary combination of efficacy outcome ($a=0$ for no efficacy observed, $a=1$ for efficacy observed) and toxicity outcome ($b=0$ for no toxicity observed, $b=1$ for toxicity observed). \\

The likelihood is then based on the categorical variable $\pi_{a,b}$ which can be in one of four states,

\begin{equation} \label{eq:LIK}
\mathcal{L}(\beta) = \prod_{i=1}^{n} \prod_{a=0}^{1} \prod_{b=0}^{1} \{\pi_{a,b}(d_{[i]},\beta)\}^{\mathbb{I}(Y=(a,b))}.
\end{equation}

Priors are elicited on $\psi$, $\beta_{E,0}$, $\beta_{E,1}$, $\beta_{T,0}$, $\beta_{T,1}$, and the likelihood (\ref{eq:LIK}) is used to update the joint posterior for all of the parameters using a Gibbs Sampler.The next dose is then chosen based on a utility function of $\pi_E$ and $\pi_T$. \\

The two main components of the design are the model (either dose-response, survival or assisting model) and the utility function. The Joint TITE-CRM and the Model-assisted method in our implementations use the same utility function, as a result of a sensitivity analysis conducted on this choice. \\

We vary the weights $w_1$ and $w_2$ in the linear utility function, and also consider the more complex utility contour method suggested by Thall and Cook \citep{Thall2004}. \\

It was found that there was negligible difference in the outcomes for any of the alternative utilities. This led to use using the simple linear utility used by Liu \& Johnson \citep{Liu2016} in our implementation of the Joint TITE-CRM, with the same weights. \\

\begin{equation}
U(\pi_E, \pi_T) = \pi_E - \omega_1 \pi_T- \omega_2 \pi_T \mathds{I}(\pi_T > \phi_T). \label{eq:utility}
\end{equation}

This uses two weights, $\omega_1$ and $\omega_2$ and a toxicity threshold $\phi_T$, above which we penalise doses that have a higher probability of DLT than this threshold. The next dose is chosen to maximise the utility out of a set of admissible doses.\\

This utility is simple to implement and to interpret. Using the same utility function for the two different designs also allows the results to be interpreted in terms of the inference used.

\subsection{Joint Model CRM} \label{sec:j_crm}
For comparison, we also consider a non-time-to-event method. For this implementation, all $\tau$ cycles must be observed from the previous cohort before the next cohort is assigned.

This is equivalent to setting $ w^{(E)}= w^{(T)}=1$ for each patient in the above Joint TITE-CRM, effectively removing the time-to-event element of the design.

\subsection{$A_T$/$A_E$ Design} \label{sec:ATAE}
We here give an overview of the $A_T$/$A_E$ design, for further details we refer the reader to the original proposal by Yuan \& Yin \citep{Yuan2009}.\\

This design fits survival models to the time to event data for both efficacy and toxicity outcomes, assuming a Weibull distribution.

\[
S_T(t|d) = \exp\{-\lambda_T t^{\alpha_T} \exp (\beta_T d) \},
\]

\[
S_E(t|d) = \exp\{-\lambda_E t^{\alpha_E} \exp (\beta_E d) \},
\]

\[
S^*_E(t|d) = 1-\pi + \pi S_E(t|d).
\]

The bivariate Time to Event data is then modelled as

\[
S(t_T,t_E|d)= \{S_T(t_T|d)^{-1/\phi} + S_E(t_E|d)^{-1/\phi} -1\}^{-\phi}.
\]
In order to compute the likelihood, define $y_T=\min (t_T,c_T)$ and $\Delta_T = \mathbb{I}(t_T \leq c_T)$ where $c_T$ is the censoring time, and define $y_E$ and $\Delta_E$ similarly for efficacy. The likelihood is then
\[
L(\theta | data_i) = L_1^{\Delta_T \Delta_E} L_2^{\Delta_T (1-\Delta_E)} L_3^{(1-\Delta_T) \Delta_E} L_4^{(1-\Delta_T) (1-\Delta_E)},
\]
where
\[
L_1=\pi \frac{\partial^2 S(y_T,y_E|d)}{\partial y_T \partial y_E},
\]

\[
L_2=-(1-\pi) \frac{\partial S_T(y_T|d)}{\partial y_T} - \pi \frac{\partial S(y_T,y_E|d)}{\partial y_T},
\]

\[
L_3= -\pi \frac{\partial S(y_T,y_E|d)}{\partial y_E},
\]

\[
L_4=(1-\pi) \partial S_T(y_T|d) + \pi S(y_T,y_E|d).
\]

$\lambda_E$, $\alpha_E$, $\beta_E$, $\lambda_T$, $\alpha_T$ and $\beta_T$ are assigned independent gamma priors and $\pi$ and $\phi$ uniform priors. The joint posterior distribution is then updated using a Gibbs Sampler.\\

The criterion used for decision making is a ratio of the area under the curve (AUC) for the survival for toxicity and efficacy:

\[
\frac{A_T}{A_E}= \frac{\alpha_T^{-1}\{\lambda_T \exp (\beta_T d)\}^{-1/\alpha_T} \Gamma \{ \alpha_T^{-1},\lambda_T \exp (\beta_T d) \tau^{\alpha_T} \}}{(1-\pi)\tau + \pi \alpha_E^{-1}\{\lambda_E \exp (\beta_E d)\}^{-1/\alpha_E} \Gamma \{ \alpha_E^{-1},\lambda_E \exp (\beta_E d) \tau^{\alpha_E} \}},
\] 

where $\Gamma (a,b)$ is the incomplete gamma function and $\tau$ is the follow up time. The dose in the admissible dose set that maximises the $A_T$/$A_E$ is chosen for the next cohort of patients.
\subsection{Model-Assisted}
The model-assisted method by Liu and Johnson \citep{Liu2016} uses a Bayesian dynamic model as follows:
\begin{align*}
y_{T,i}|d=d_j &\sim \mbox{Bern} (p_{T,j}) \\
p_{T,j} &= p_{T,j-1} + ( 1-p_{T,j-1} )\beta_{T,j} , \hspace{10pt} j=2,\ldots J  \\
p_{T,1} &= \beta_{T,1}  \\
\beta_{T,j} &\sim \mbox{Beta} (a_{T,j},b_{T,j}) \hspace{10pt} j=1,\ldots J.
\end{align*}
The subscript $T$ indicates toxicity, with the equivalent for efficacy labelled $E$. The specification of the $p_{T,j}$ ensures monotonicity, since there is no dose-response model.\\

The likelihood is given as 
\[
L(\mathbf{y} | \mathbf{\beta_T},\mathbf{\beta_E}) = \prod_{k=\{E,T\}} \prod_{i=1}^{n} \left\{ 1- \prod_{r=1}^{j[i]} (1-\beta_{k,r}) \right\}^{y_{k,j}} \prod_{r=1}^{j[i]} (1 - \beta_{k,r})^{1-y_{k,j}},
\]
which is used to update the posterior distribution. The same utility function as is used in the Joint TITE-CRM (Equation~\ref{eq:utility}) is used to assign the next dose.

\subsection{Rules} \label{sec:rules}
\subsubsection{Admissible Dose Set} \label{sec:admis}
The admissible dose set is calculated as all doses satisfying the following criteria:
\[
\mathbb{P} (\pi_T < \pi_T^*) > q_T 
\hspace{15pt} \mbox{\&} \hspace{15pt} 
\mathbb{P} (\pi_E > \pi_E^*) > q_E.
\]
where $\pi_T^*$ and $\pi_E^*$ are the target probabilities for efficacy and toxicity in the entire follow-up $\tau$.\\

These constraints ensure that for both designs, the escalation proceeds to a promising dose that is considered neither futile or unsafe.
\subsubsection{Stopping \& Enforcement Rules} \label{sec:stop_enf_rules}
As well as ensuring any assigned dose is admissible, we also apply a set of enforcement and stopping rules.
For any given dose $d_j$, $p_{1,d_j}$ is the $P(DLT)$ in the first cycle, $p_{\tau,d_j}$ is the $P(DLT)$ in the full follow up of $\tau$.\\
Enforcement rules are to ensure the safety in the first cycle of treatment.
\textbf{\textit{Enforcement Rules:}}
\begin{enumerate}
\item \textbf{Hard Safety}: This rule ensures that if there is a very high probability that the toxicity of an experimented dose exceeds 0.3 in the first cycle, then that dose and all above are excluded from any further experimentation.(i.e. dose $d_j$ and all above are excluded when $P(p_{1,d_j}>0.3)>\zeta$ for some threshold $\zeta$). In this implementation we use a threshold for excessive toxicity of $\zeta =0.95$, with a $Beta(1,1)$ prior for Binomial responses. This is equivalent to if there are at least 3 DLT responses out of 3 patients, at least 4 DLT responses out of 6 patients, or at least 5 DLT responses out of 9 patients, then all dose assignments must be strictly lower than that dose for the rest of the study. If the lowest dose is excluded then the trial stops with no dose recommendation made.
\item \textbf{K-fold Skipping Doses}: The next dose assignment must be no more than a 2-fold-rise in the value of the highest experimented dose so far.
\end{enumerate}

Stopping rules are to define when we may stop the trial, either for futility/safety issues or because we have a level of certainty about the estimated OBD.

\textbf{\textit{Stopping Rules:}}
\begin{enumerate}
\item \textbf{No Admissible Doses}: If no doses satisfy the two constraints in Section~\ref{sec:admis}, then the trial is stopped, either for futility or safety.
\item \textbf{Lowest Dose Deemed Unsafe}: If $P(p_{1,d_1}>30\%)>0.80$ and at least one cohort of patients has been assigned to dose $d_1$, the trial is stopped. \label{rule:low_unsafe}
\item \textbf{Highest Dose Deemed Very Safe}: If $P(p_{1,d_J}\leq 30\%)>0.80$ and at least one cohort of patients has been assigned to dose $d_J$, the trial is stopped. \label{rule:hi_safe}
\item \textbf{Sufficient Information}: If a dose is recommended for the next cohort on which $C_{suff}$ cohorts have already been assigned in the escalation, excluding backfilled cohorts, the trial is stopped. \label{rule:suff}
\item \textbf{Precision}: If the safety and efficacy profile are both estimated precisely enough, the trial is stopped. This precision is defined as $CV(MTD)<30\%$ and $CV(d_{[\pi^*_E]})<30\%$, with the coefficient of variation calculated as an adjusted median absolute deviation divided by the median and $d_{[\pi^*_E]}$ is the dose associated with the target efficacy. Both of these must be satisfied before the stopping rule is enforced. This stopping rule is only used once at least $C_{suff}$ cohorts of patients have had at least one cycle of treatment in the escalation, on any dose. This rule is not applicable for the model-assisted method. \label{rule:prec}
\item \textbf{Hard Safety}: If the lowest dose is considered unsafe according to the hard safety enforcement rule, the trial is stopped.
\item \textbf{Maximum Patients}: If the maximum number of patients ($n=n_{\mbox{max}}$) have been recruited, the trial is stopped.
\end{enumerate}

We consider two settings; a realistic setting where all enforcement and stopping rules are applied, and a theoretical setting where stopping rules \ref{rule:low_unsafe}, \ref{rule:hi_safe}, \ref{rule:suff}, \ref{rule:prec} are not applied. This second setting is to investigate the behaviour of the designs without restrictions.

\section{Simulations} \label{sec:sims}
In order to compare the joint TITE-CRM to both the $A_T/A_E$ design and the model-assisted design, we conduct simulation studies in a range of scenarios.

\subsection{Setting} \label{sec:setting}
We consider the setting of six doses to investigate: 1.5MBq , 2.5MBq , 3.5MBq , 4.5MBq , 6.0MBq , 7.0MBq, with a follow-up period of $\tau=3$ cycles. Each cycle lasts 6 weeks, although for simplicity of time-to-event outcomes, we measure the time to event in units of cycles. The variables for the rules are $n_{max}=60$ and $C_{suff}=30$.
\\

We use a combination of five safety scenarios and four efficacy scenarios, giving 20 scenarios in total. We then extend to consider different efficacy patterns in Section~\ref{sec:eff_time_trends}. The five safety scenarios represent four scenarios where there is at least one safe dose and one where no doses are safe. We give the probability of DLT in cycle 1 and in the full follow up period of 3 cycles, with $\pi_T^*=0.391$. It is assumed that the probability of DLT decreases by a factor of 1/3 in subsequent cycles, conditional on survival through the previous cycles. In T.1, dose levels 4 and above are unsafe; in T.2, the highest dose is unsafe; in T.3, all doses are unsafe; in T.4, dose levels 2 and above are unsafe; and in T.5 all doses are well below target safety. The four efficacy scenarios are described by the probability of efficacy in the whole follow up period, with the lower bound on target efficacy as $\pi_T^*=0.2$. In E.1, dose levels 2 and above are efficacious, with efficacy increasing with dose; in E.2, all dose levels are efficacious with a plateau reached at dose level 3; in E.3, no dose levels are efficacious; and in E.4, dose levels 4 and above are efficacious.\\

In the main implementations, it is assumed the probability of efficacy in the total follow up is three times the probability of efficacy in the first cycle. In Section~\ref{sec:eff_time_trends}, we alter this assumption.\\

Scenarios are referred to as Ex.Ty where x is the efficacy scenario, y is the safety scenario. Table~\ref{tab:scens} gives the individual safety and efficacy scenarios, while Table~\ref{tab:scens_ut} gives the utility and $AT/AE$ values for each of the 20 combination scenarios.
\begin{table}[ht]
\centering
\begin{tabular}{lllllll}
   \hline
\hline
Safety & 1.5MBq & 2.5MBq & 3.5MBq & 4.5MBq & 6.0MBq & 7.0MBq \\ 
   \hline
T1 (cycle 1)& 0.100 & 0.200 & 0.300 & 0.400 & 0.500 & 0.600 \\ 
T1 (full follow up)& 0.140 & 0.270 & 0.391 & 0.503 & 0.606 & 0.701 \\ 
  T2 (cycle 1)& 0.100 & 0.130 & 0.160 & 0.200 & 0.250 & 0.400 \\ 
 T2 (full follow up) & 0.140 & 0.180 & 0.219 & 0.270 & 0.332 & 0.503 \\ 
  T3 (cycle 1)& 0.400 & 0.450 & 0.500 & 0.550 & 0.600 & 0.650 \\ 
 T3 (full follow up) & 0.503 & 0.556 & 0.606 & 0.655 & 0.701 & 0.746 \\ 
  T4 (cycle 1)& 0.300 & 0.400 & 0.450 & 0.500 & 0.550 & 0.600 \\ 
  T4 (full follow up)& 0.391 & 0.503 & 0.556 & 0.606 & 0.655 & 0.701 \\ 
    T5 (cycle 1)&0.100 &0.120 &0.140 &0.160 &0.180 &0.200  \\ 
  T5 (full follow up)&0.140 &0.166 &0.193 &0.219 &0.245& 0.270 \\ 
   \hline
\hline
Efficacy & 1.5MBq & 2.5MBq & 3.5MBq & 4.5MBq & 6.0MBq & 7.0MBq \\ 
   \hline
E1 & 0.200 & 0.300 & 0.400 & 0.500 & 0.600 & 0.700 \\ 
  E2 & 0.300 & 0.400 & 0.500 & 0.500 & 0.500 & 0.500 \\ 
  E3 & 0.100 & 0.120 & 0.140 & 0.160 & 0.180 & 0.200 \\ 
  E4 & 0.100 & 0.150 & 0.200 & 0.300 & 0.500 & 0.700 \\ 
   \hline
\end{tabular}
\caption{Scenario definitions for efficacy and toxicity.}
\label{tab:scens}
\end{table}

\begin{table}[ht]
\centering
\begin{tabular}{rrrrrrr} \hline
Scenario  & 1.5MBq & 2.5MBq & 3.5MBq & 4.5MBq & 6.0MBq & 7.0MBq \\ \hline  \hline
 E1.T1 Utility & \textit{0.15} & \textit{0.21} & \textbf{0.27 }& -0.22 & -0.26 & -0.30 \\ 
  E1.T1 $AT/AE$ & \textit{0.99} & \textit{0.93} & \textbf{0.86} & 0.79 & 0.71 & 0.61 \\ 
   \hline
E1.T2 Utility & \textit{0.15} & \textit{0.24} & \textit{0.33} & \textit{0.41} & \textbf{0.49} & -0.02 \\ 
  E1.T2 $AT/AE$ & \textit{0.99} & \textit{1.02} & \textit{1.05} & \textit{1.06} & \textbf{1.07} & 0.92 \\ 
   \hline
E1.T3 Utility & -0.52 & -0.49 & -0.46 & -0.43 & -0.40 & -0.36 \\ 
  E1.T3 $AT/AE$ & 0.65 & 0.63 & 0.61 & 0.59 & 0.57 & 0.54 \\ 
   \hline
E1.T4 Utility & \textit{0.07} & -0.42 & -0.39 & -0.36 & -0.33 & -0.30 \\ 
  E1.T4 $AT/AE$ & \textit{0.76} & 0.69 & 0.67 & 0.66 & 0.64 & 0.61 \\ 
   \hline
E1.T5 Utility & \textit{0.15} & \textit{0.25} & \textit{0.34} & \textit{0.43} & \textit{0.52} & \textit{0.61} \\ 
  E1.T5 $AT/AE$ & \textit{0.99} & \textit{1.03} & \textit{1.07} & \textit{1.12} & \textit{1.17} & \textit{1.23} \\ 
   \hline
E2.T1 Utility & \textit{0.25} & \textit{0.31} & \textbf{0.37} & -0.22 & -0.36 & -0.50 \\ 
  E2.T1 $AT/AE$ & \textit{1.05} & \textit{0.99} & \textbf{0.92} & 0.79 & 0.66 & 0.53 \\ 
   \hline
E2.T2 Utility & \textit{0.25} & \textit{0.34} &\textbf{ 0.43} & 0.41 & 0.39 & -0.22 \\ 
  E2.T2 $AT/AE$ & \textit{1.05} & \textit{1.08} & \textbf{1.12} & 1.06 & 0.99 & 0.79 \\ 
   \hline
E2.T3 Utility & -0.42 & -0.39 & -0.36 & -0.43 & -0.50 & -0.56 \\ 
  E2.T3 $AT/AE$ & 0.69 & 0.67 & 0.66 & 0.59 & 0.53 & 0.46 \\ 
   \hline
E2.T4 Utility & \textbf{0.17} & -0.32 & -0.29 & -0.36 & -0.43 & -0.50 \\ 
  E2.T4 $AT/AE$ & \textbf{0.81} & 0.74 & 0.72 & 0.66 & 0.59 & 0.53 \\ 
   \hline
E2.T5 Utility & \textit{0.25} & \textit{0.35} & \textit{0.44 }& \textit{0.43} & \textit{0.42} & \textit{0.41} \\ 
  E2.T5 $AT/AE$ & \textit{1.05} & \textit{1.10} & \textit{1.14} & \textit{1.12} & \textit{1.09} & \textit{1.06} \\ 
   \hline
E3.T1 Utility & 0.05 & 0.03 & 0.01 & -0.56 & -0.68 & -0.80 \\ 
  E3.T1 $AT/AE$ & 0.94 & 0.84 & 0.74 & 0.64 & 0.54 & 0.44 \\ 
   \hline
E3.T2 Utility & 0.05 & 0.06 & 0.07 & 0.07 & 0.07 & -0.52 \\ 
  E3.T2 $AT/AE$ & 0.94 & 0.92 & 0.89 & 0.86 & 0.81 & 0.65 \\ 
   \hline
E3.T3 Utility & -0.62 & -0.67 & -0.72 & -0.77 & -0.82 & -0.86 \\ 
  E3.T3 $AT/AE$ & 0.62 & 0.57 & 0.53 & 0.48 & 0.43 & 0.38 \\ 
   \hline
E3.T4 Utility & -0.03 & -0.60 & -0.65 & -0.70 & -0.75 & -0.80 \\ 
  E3.T4 $AT/AE$ & 0.72 & 0.63 & 0.58 & 0.53 & 0.48 & 0.44 \\ 
   \hline
E3.T5 Utility & 0.05 & 0.07 & 0.08 & 0.09 & 0.10 & \textit{0.11} \\ 
  E3.T5 $AT/AE$ & 0.94 & 0.93 & 0.92 & 0.90 & 0.89 & \textit{0.88} \\ 
   \hline
E4.T1 Utility & 0.05 & 0.06 & \textit{0.07} & -0.42 & -0.36 & -0.30 \\ 
  E4.T1 $AT/AE$ & 0.94 & 0.85 & \textit{0.76} & 0.69 & 0.66 & 0.61 \\ 
   \hline
E4.T2 Utility & 0.05 & 0.09 & \textit{0.13} & \textit{0.21} & \textbf{0.39} & -0.02 \\ 
  E4.T2 $AT/AE$ & 0.94 & 0.93 & \textit{0.92} & \textit{0.93} & \textbf{0.99} & 0.92 \\ 
   \hline
E4.T3 Utility & -0.62 & -0.64 & -0.66 & -0.63 & -0.50 & -0.36 \\ 
  E4.T3 $AT/AE$ & 0.62 & 0.58 & 0.54 & 0.52 & 0.53 & 0.54 \\ 
   \hline
E4.T4 Utility & -0.03 & -0.57 & -0.59 & -0.56 & -0.43 & -0.30 \\ 
  E4.T4 $AT/AE$ & 0.72 & 0.64 & 0.60 & 0.58 & 0.59 & 0.61 \\ 
   \hline
E4.T5 Utility & 0.05 & 0.10 & \textit{0.14} & \textit{0.23} &\textit{ 0.42} & \textit{0.61} \\ 
  E4.T5 $AT/AE$ & 0.94 & 0.94 & \textit{0.95} & \textit{0.98} & \textit{1.09} & \textit{1.23} \\ 
   \hline
\end{tabular}
\caption{Utility for the twenty scenarios, with OBD highlighted in \textbf{bold} and acceptable doses highlighted in \textit{italics}. Note that the $AT/AE$ criteria has no penalty for exceeding a safety threshold and hence may give higher values for unsafe doses.}
 \label{tab:scens_ut}
\end{table}

In Table~\ref{tab:scens_ut}, it can be seen that the utility criterion and $AT/AE$ criterion do not always agree on the OBD. For example in scenarios E1.T1 and E2.T1, the 3.5MBq dose is the OBD according to the utility criterion, and the 1.5MBq dose is the OBD according to the $AT/AE$ criterion. We have chosen to designate the 3.5MBq dose as the true OBD as this is more realistic, with a P(DLT) equal to target and a higher P(efficacy) than the lower dose. This does however highlight an issue concerning the $AT/AE$ criteria, in that in maximizing the ratio between the two areas under the curves, it does not target any specific safety level. Hence when ratios are similar across doses, even when values are not, there may be difficulties in selection of the optimal dose.

\subsection{Data Generation} \label{sec:data_gen}
To generate the event times for toxicities and efficacies, we use a model that is not based on the assumptions of any of the methods. We use a bivariate log-normal distribution for survival times, with parameters matched to the first cycle and full follow up probabilities of efficacy and toxicity, with a correlation parameter of -1/2. This indicates that there is a negative association between the event times. This is a simple yet effective mechanism to generate data that allows for easy specification of probabilities at dose levels that do not themselves follow any specific parametric dose-response model.\\
\[
\begin{pmatrix}
t_T \\
t_E
\end{pmatrix} \sim Lognormal_2 \left( \begin{pmatrix}
\mu_T\\
\mu_E
\end{pmatrix} , \begin{pmatrix}
\sigma^2_T & -\frac{\sigma_T \sigma_E}{2}\\
-\frac{\sigma_T \sigma_E}{2} & \sigma^2_E
\end{pmatrix} \right).
\]

\subsection{Prior distributions} \label{sec:prior}
Since all methods are Bayesian, we must specify prior distributions for the parameters of the models.\\
\subsubsection{Joint Model TITE-CRM \& Joint Model CRM}
Here we use priors
\[
\begin{pmatrix}
\beta_0 \\
\log (\beta_1)
\end{pmatrix} \sim N_2 \left( \begin{pmatrix}
c_1\\
c_2
\end{pmatrix} , \begin{pmatrix}
v_1 & 0 \\
0 & v_2 
\end{pmatrix} \right).
\]
for both toxicity and efficacy. Values of hyper-parameters used are:
$c_{1,T}=\log (1/16) $, $c_{2,T}=\log (1/4)$, $v_{1,T}=1$, $v_{2,T}=2$ and $c_{1,E}= -3 $, $c_{2,E}=-0.2$, $v_{1,E}=1$, $v_{2,E}=1$. The values for toxicity are calibrated over a range of scenarios. The values for efficacy are chosen so that all doses have prior mean $(\pi_E)>$ 0.3 and this increases with dose, with average effective prior sample size of one patient per dose level. We used a vague prior for $\phi \sim N(0,100)$.

\subsubsection{$A_T$/$A_E$ Design}
Here, we follow the priors implemented by Yuan \& Yin \citep{Yuan2009}, with some minor modification to fit our setting. $\lambda_E$, $\alpha_E$, $\beta_E$, $\lambda_T$, $\alpha_T$ and $\beta_T$ are assigned independent Gamma priors with shape and rate parameters of 0.1. These are constrained to be less than 3, 4, 1, 2, 3 and 1 respectively. The modification is of the upper bound of the $\beta$ to account for the differing values of doses in our implementation to the original implementation. The uniform priors for $\pi$ and $\phi$ are $\pi\sim U(0.6,1)$ and $\phi\sim U(0,5)$.
\subsubsection{Model-Assisted}
Here we use the priors suggested by Liu \& Johnson \cite{Liu2016} of $\mathbf{a_E}=(0.2,0.3,0.4,0.45,0.5,0.6)$, $\mathbf{b_E}=(0.95,0.9,0.8,0.75,0.7,0.65)$, $\mathbf{a_T}=(0.05,0.1,0.2,0.25,0.3,0.35)$ and $\mathbf{b_T}=(0.95,0.9,0.8,0.75,0.7,0.65)$.

\subsection{Results} \label{sec:results}
\begin{figure}
      \includegraphics[width=1\linewidth]{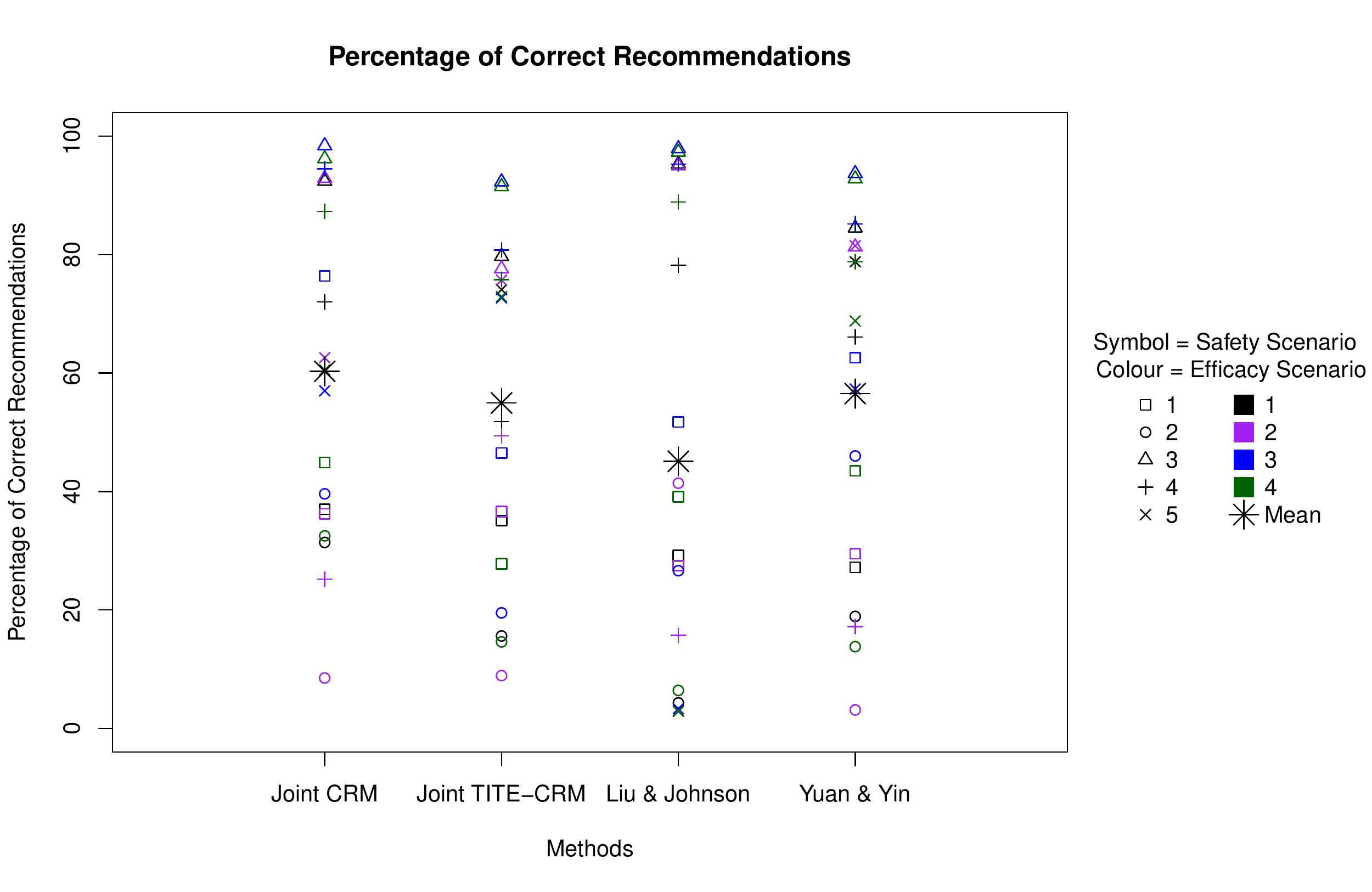}
     \caption{Proportion of correct selections across scenarios}   \label{fig:correct}
  \end{figure}

We conducted 1000 simulations in each of the 20 scenarios. Figure~\ref{fig:correct} shows the percentage of simulations that select the true OBD if there is one, or the trial is stopped correctly when no true OBD exists in the experimental dose set. The most noticeable feature is that the model-assisted method gives a lower proportion of the correct recommendations than the two model-based methods. In scenarios where all doses are too safe, the model-assisted method performs poorly, a reflection on the cautious escalation so often apparent in model-assisted designs. The overall performance of the two model based methods is similar, although with some differences in individual scenarios. For example the performance in E2.T4, where the lowest dose is the OBD, the Joint TITE-CRM method vastly outperforms the $AT/AE$ design, which too often stops early for no admissible doses. \\

\begin{figure}
      \includegraphics[width=1\linewidth]{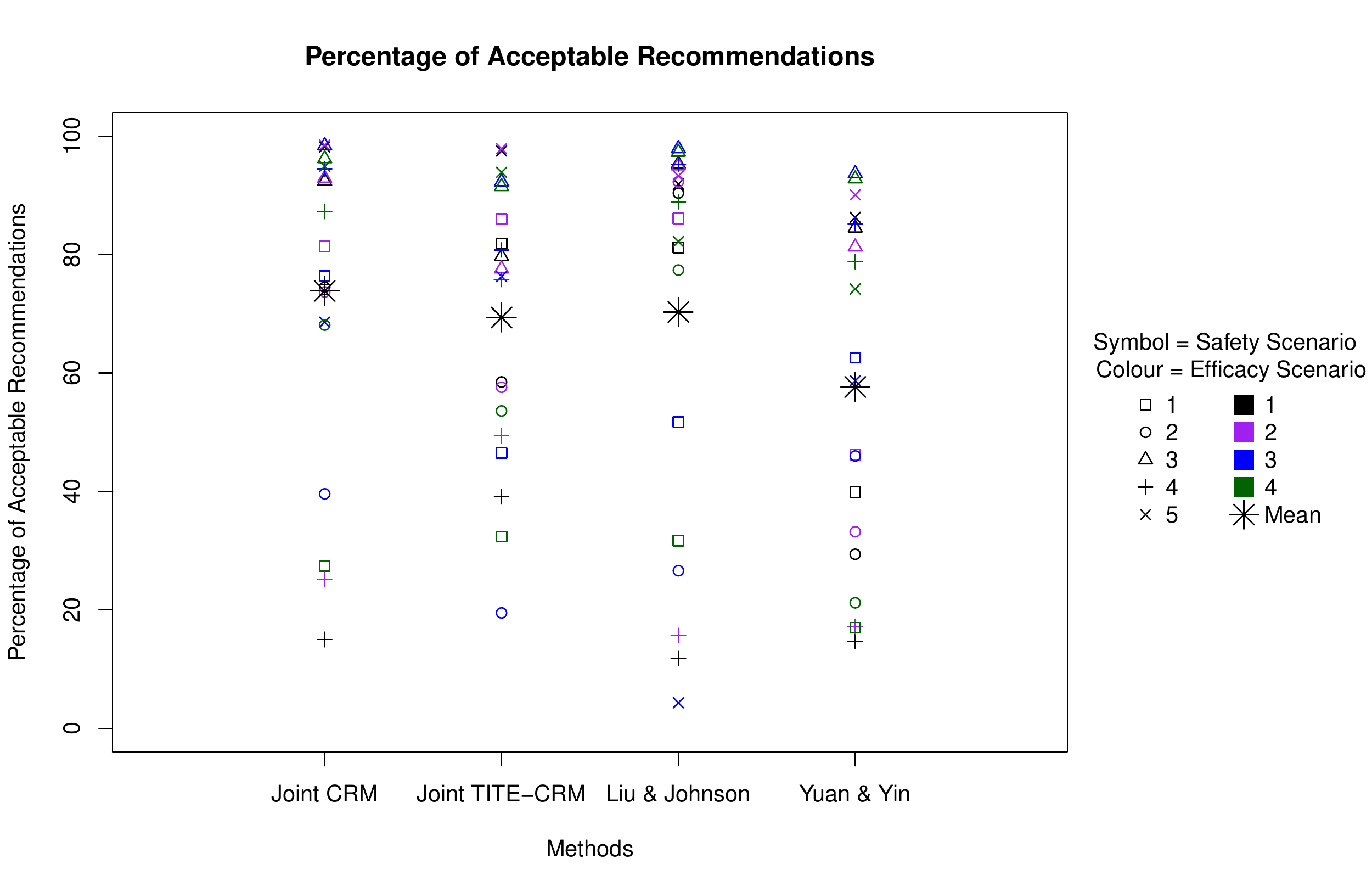}
     \caption{Proportion of acceptable selections across scenarios}   \label{fig:acceptable}
  \end{figure}

Interestingly, the percentage of acceptable selections does not follow such a similar pattern. Figure~\ref{fig:acceptable} displays these results. Here we define an `acceptable' selection as one that is both safe and efficacious, regardless if it the best in terms of utility criterion. The Joint TITE-CRM is the best performing overall. The Model-assisted method performs poorly for example in E3.T5, where the acceptable selections are either the highest dose or stopping because the highest dose is too safe.\\

\begin{figure}
      \includegraphics[width=1\linewidth]{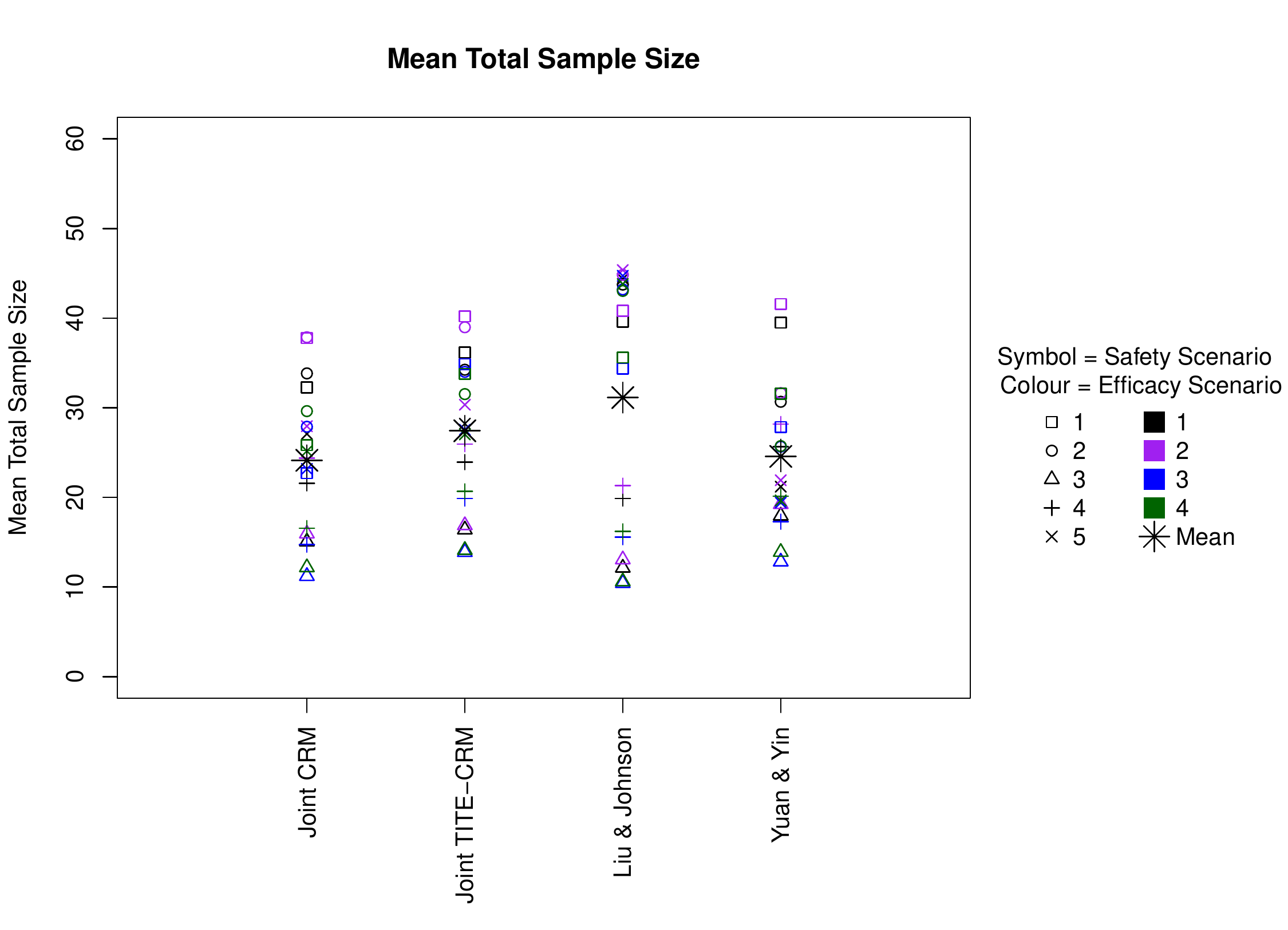}
     \caption{Mean sample size across scenarios}   \label{fig:ssize}
  \end{figure}

In terms of sample size, the model assisted method has a higher sample sample size on average across scenarios, as shown in Figure~\ref{fig:ssize}. The two model-based methods have very similar sample sizes.\\

  \begin{figure}
      \includegraphics[width=1\linewidth]{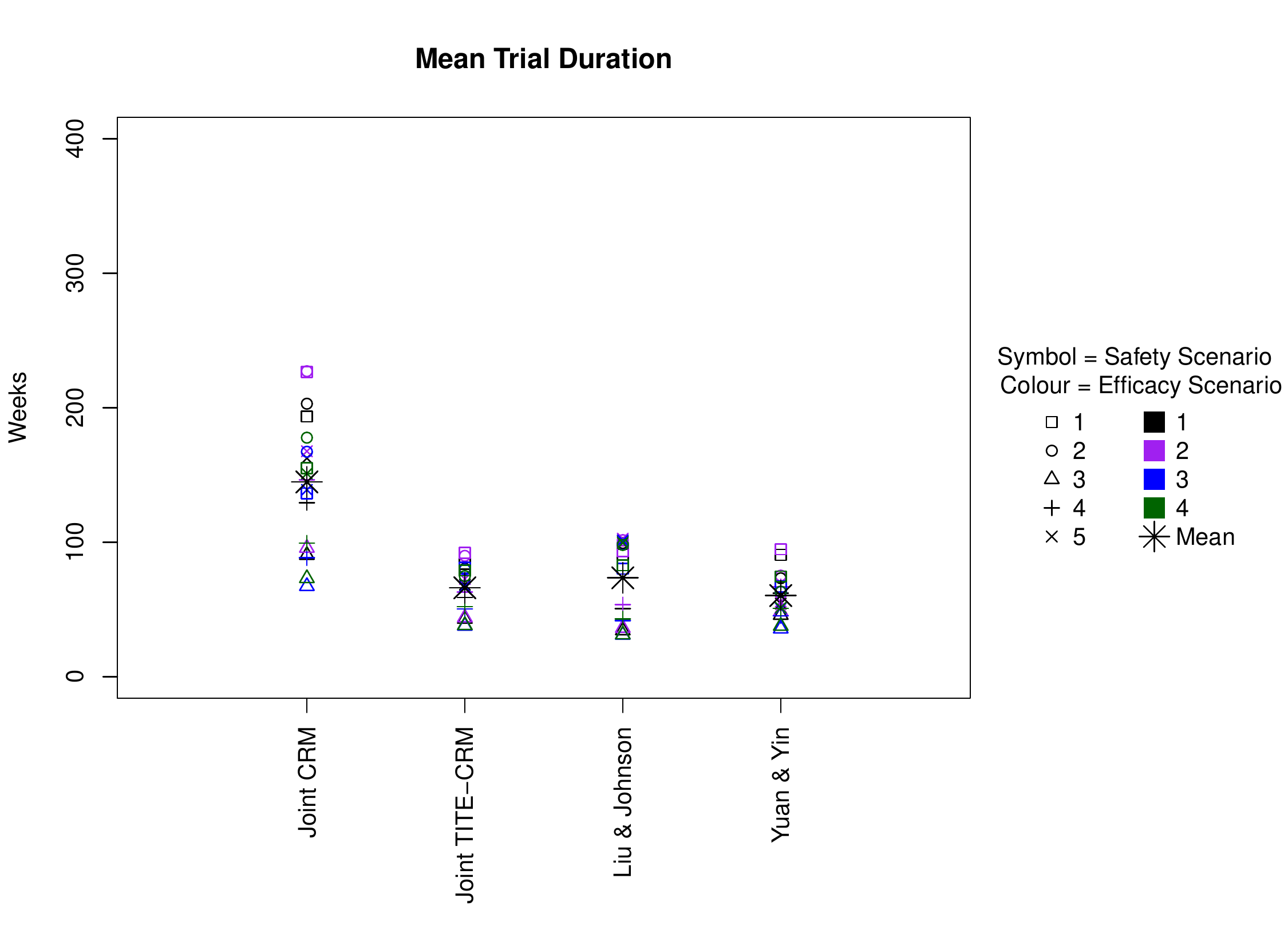}
     \caption{Mean trial duration across scenarios}   \label{fig:dur}
  \end{figure}
However, although the patterns are similar between the two model-based methods in terms of total sample sizes, correct and acceptable selections, there is a difference in the number of patients treated at unsafe doses, with the $AT/AE$ method assigning more patients to unsafe doses than both other methods in almost all scenarios. The model assisted method has much fewer patients assigned to unsafe doses, again due to the cautious escalation behaviour of model-assisted methods. \\

Although the performance of the Joint CRM is similar to that of the Joint TITE-CRM, with slight improvement, this is at the cost of a much greater average trial length. Figure~\ref{fig:dur} indicated this relationship, with an average trial duration of more than twice that of the Joint TITE-CRM. This is a sure indication of the need for trial designs that can use partial information on cohorts effectively.

  \begin{figure}
      \includegraphics[width=1\linewidth]{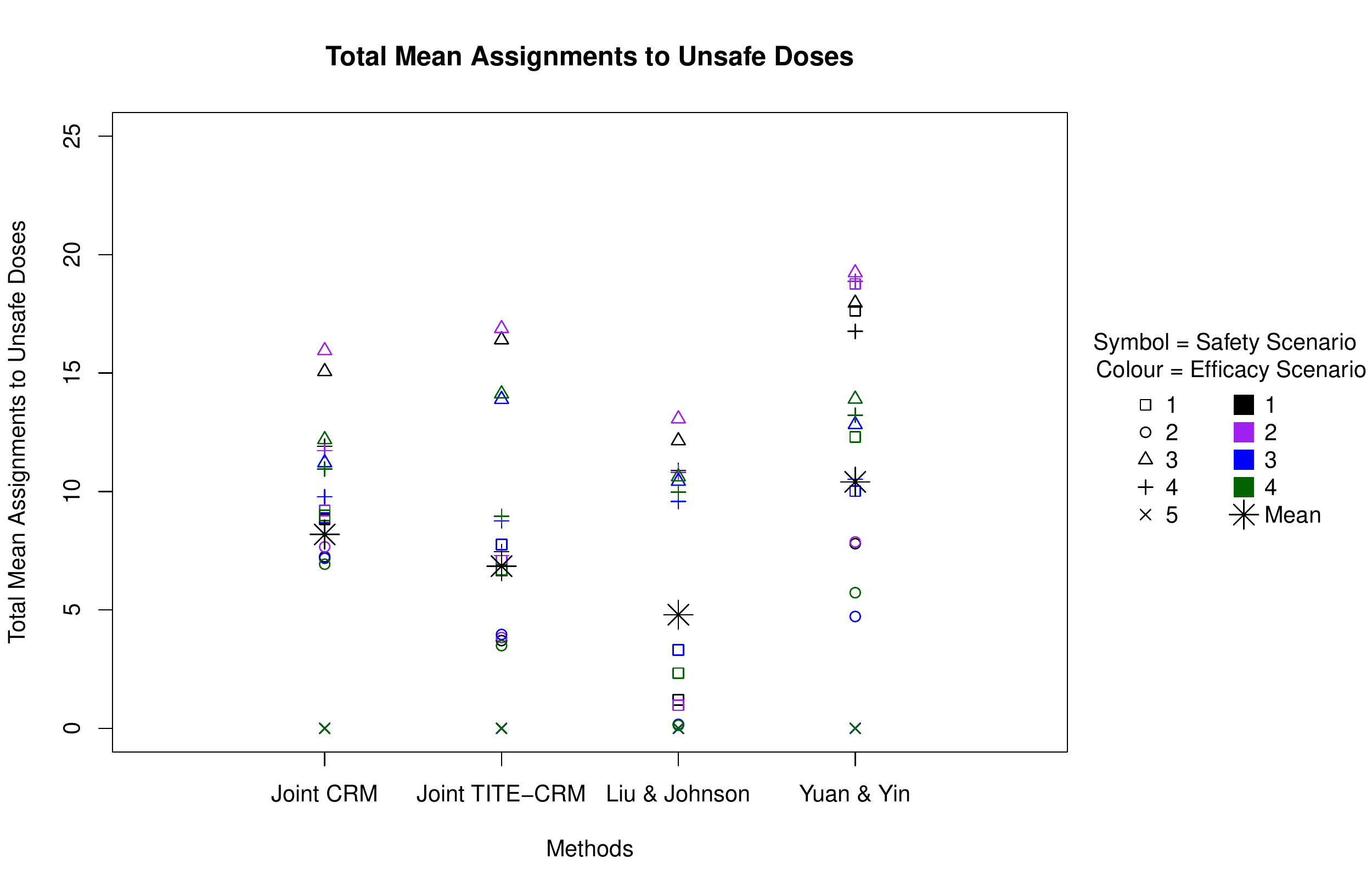}
     \caption{Number of patients assigned to unsafe doses across scenarios}   \label{fig:unsafe}
  \end{figure}

The overall performance of the three methods in the setting of realistic stopping and enforcement rules can be summarised as follows. The cautious escalation of the model-assisted method gives rise to a poorer performance in terms of selections of correct and acceptable doses, with larger sample sizes and fewer patients exposed to unsafe doses. The two model based designs perform better, with similar percentages of correct and acceptable recommendations, but with the $AT/AE$ design assigning more patients to unsafe doses, showing a more aggressive escalation that is not rewarded with better selection performance. Additionally, the $AT/AE$ design is substantially more complex and computationally intensive than both other methods, a cost that does not increase the performance.\\

When the stopping rules are relaxed to investigate the operating characteristics of the designs more closely, note that since stopping rule \ref{rule:hi_safe} is no longer enforced, there is no longer a correct selection in safety scenario 5 and so all methods give 0\% correct selection. We see that in most scenarios where there is a true OBD (e.g. E1.T2, E2.T4, E4.T2) the Joint TITE-CRM gives a better performance of correct selections. When there are no admissible doses (e.g. E3.T1, E3.T2) the Joint TITE-CRM performs worse than the other two time-to-event methods. 

It is noticeable that when considering acceptable selections, the Joint TITE-CRM performs the best out of the selection in  most scenarios. In the absence of the early stopping rules, this provides evidence that the Joint TITE-CRM is benefiting from the model-based inference that the model-assisted method lacks. However, the $AT/AE$ method also performs worse than the Joint TITE-CRM. This may be due to the deviations from assumptions on the survival model used in the inference and data generation.

The sample size is of course considerably larger when the stopping rules are relaxed, with many scenarios reaching nearly the maximum sample size on average. The Joint TITE-CRM has a larger sample size on average, which tallies up to the observation that this design had fewer correct stoppings for no admissible doses. This suggests that in general, this method is more willing to label a dose as admissible than the other methods.\\

The mean trial duration is also drastically increased for all methods, with the Joint CRM having an average duration of up to seven years, compared to the other methods giving less than three years.\\

With the relaxed stopping rules, the $AT/AE$ method once again has a large number of patients assigned to unsafe doses, indicating a more aggressive escalation. The model-assisted method has much fewer patients assigned to unsafe doses, due to the more cautious escalation.

  \begin{figure}
      \includegraphics[width=1\linewidth]{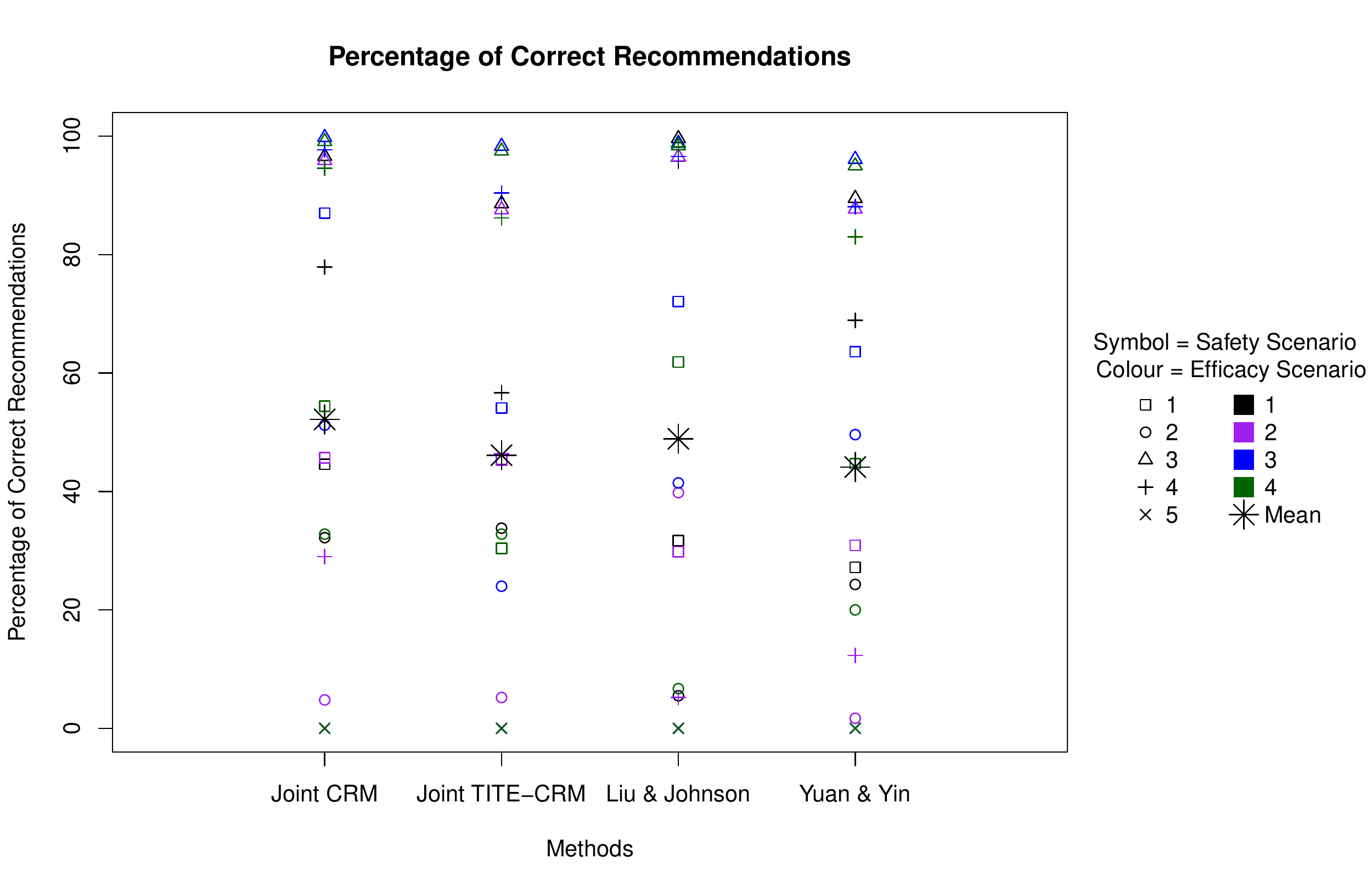}
     \caption{Proportion of correct selections across scenarios, lesser stopping}   \label{fig:correctnoSTOP}
  \end{figure}

  \begin{figure}
      \includegraphics[width=1\linewidth]{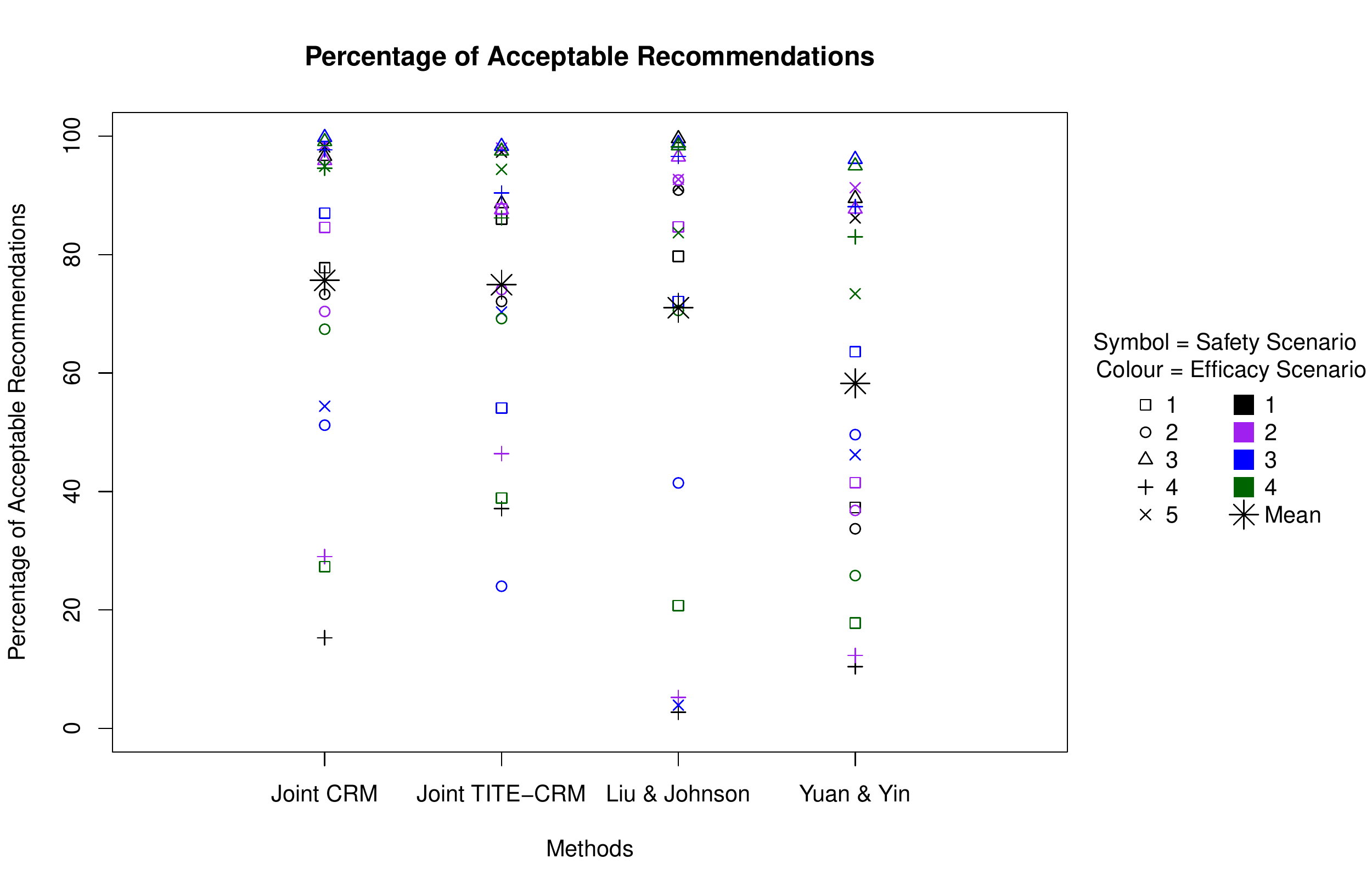}
     \caption{Proportion of acceptable selections across scenarios lesser stopping}   \label{fig:acceptablenoSTOP}
  \end{figure}

  \begin{figure}
      \includegraphics[width=1\linewidth]{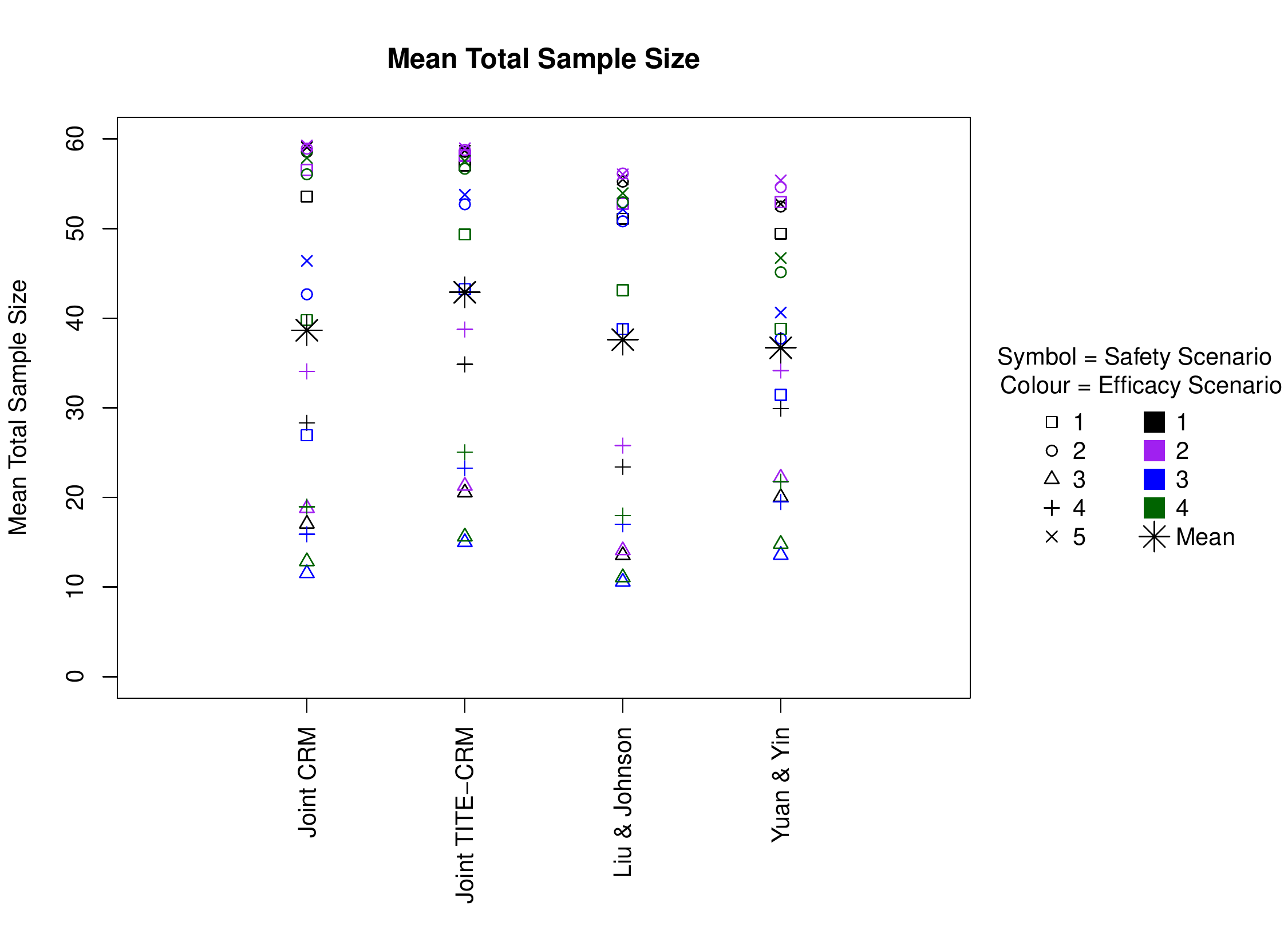}
     \caption{Mean sample size across scenarios lesser stopping}   \label{fig:ssizenoSTOP}
  \end{figure}

  \begin{figure}
      \includegraphics[width=1\linewidth]{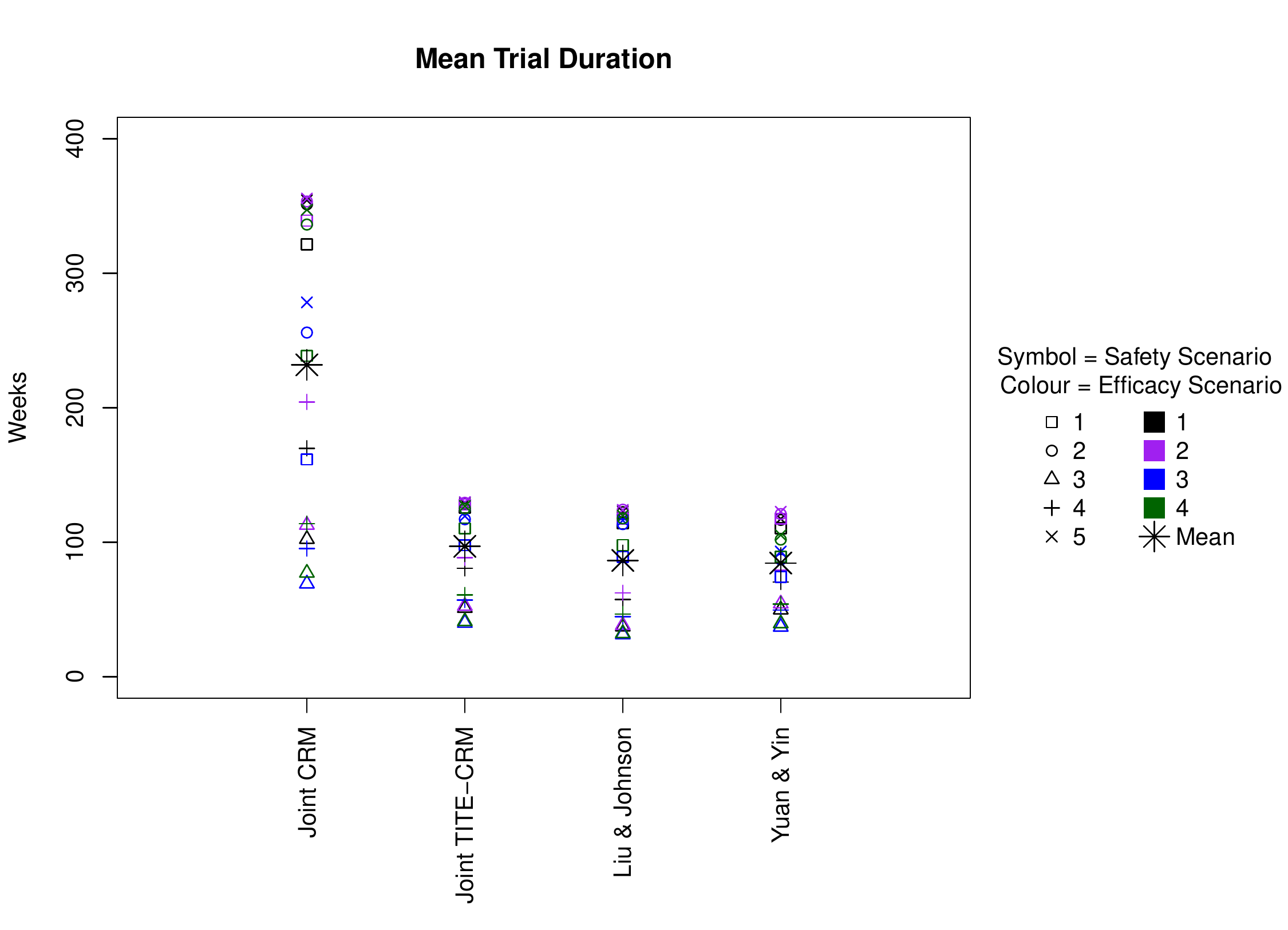}
     \caption{Mean trial duration across scenarios lesser stopping}   \label{fig:durnoSTOP}
  \end{figure}

  \begin{figure}
      \includegraphics[width=1\linewidth]{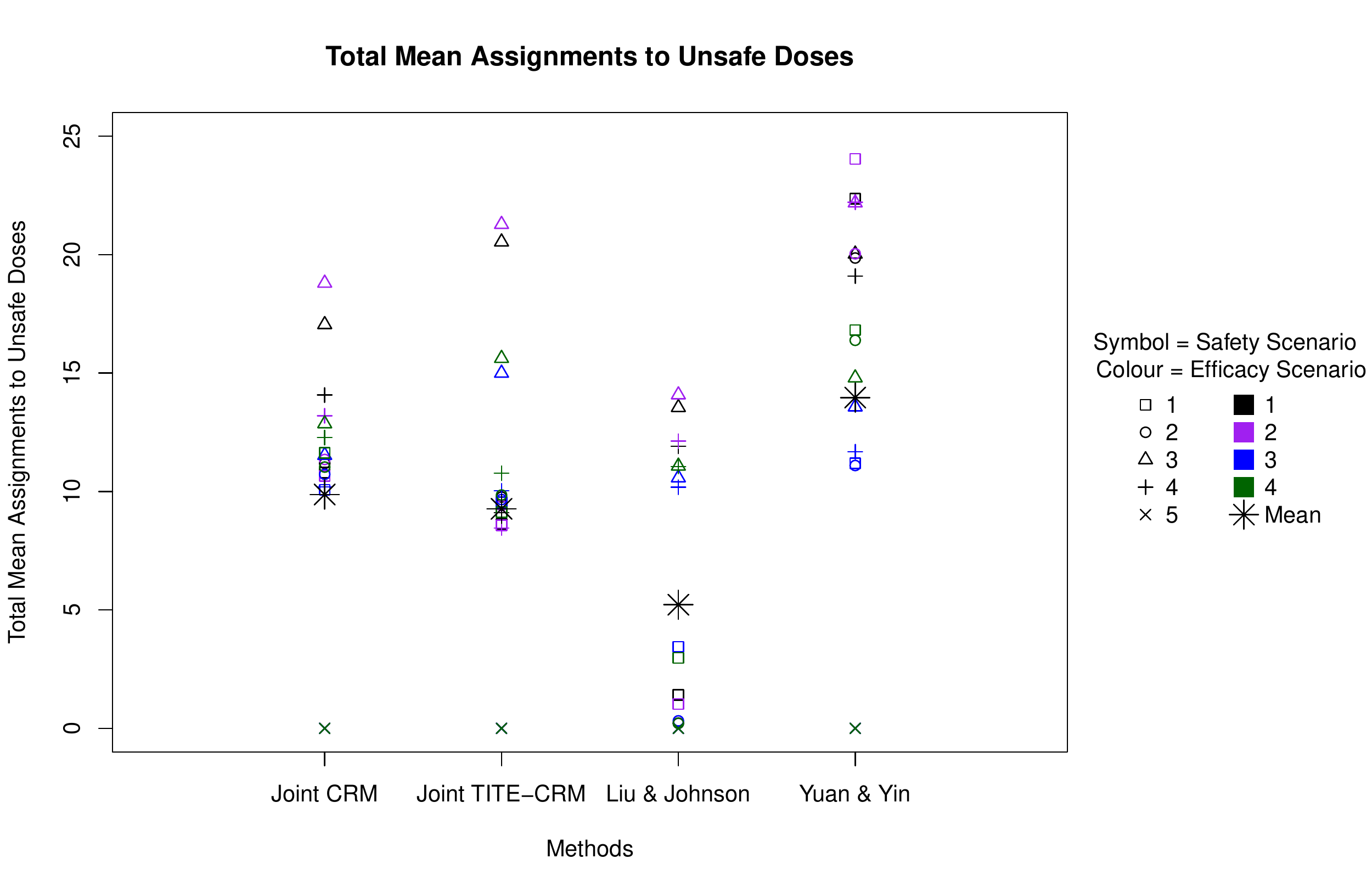}
     \caption{Number of patients assigned to unsafe doses across scenarios lesser stopping}   \label{fig:unsafenoSTOP}
  \end{figure}

\subsection{Efficacy Time Trend} \label{sec:eff_time_trends}
In the main implementations, the data generation of the efficacy times assumed that the probability of efficacy in the first cycle is one third of the probability of efficacy in the entire follow up of three cycles. However, it may be the case that this is actually higher or lower than one third.\\

To investigate the effect of efficacy occurring at various points in the follow up, we consider three efficacy patterns. In efficacy pattern 1, the probability of efficacy in cycle 1 is 1/3 of the total probability; in efficacy pattern 2, it is 1/6 and in efficacy pattern 3 in is 1/2. This is to investigate the effect of varying time trends in efficacy.\\

It is possible that this change in efficacy time trend will affect the performance of the three methods that us the time-to-event outcomes, and so we investigate the difference that this can make in a selection of scenarios.

  \begin{figure}
      \includegraphics[width=1\linewidth]{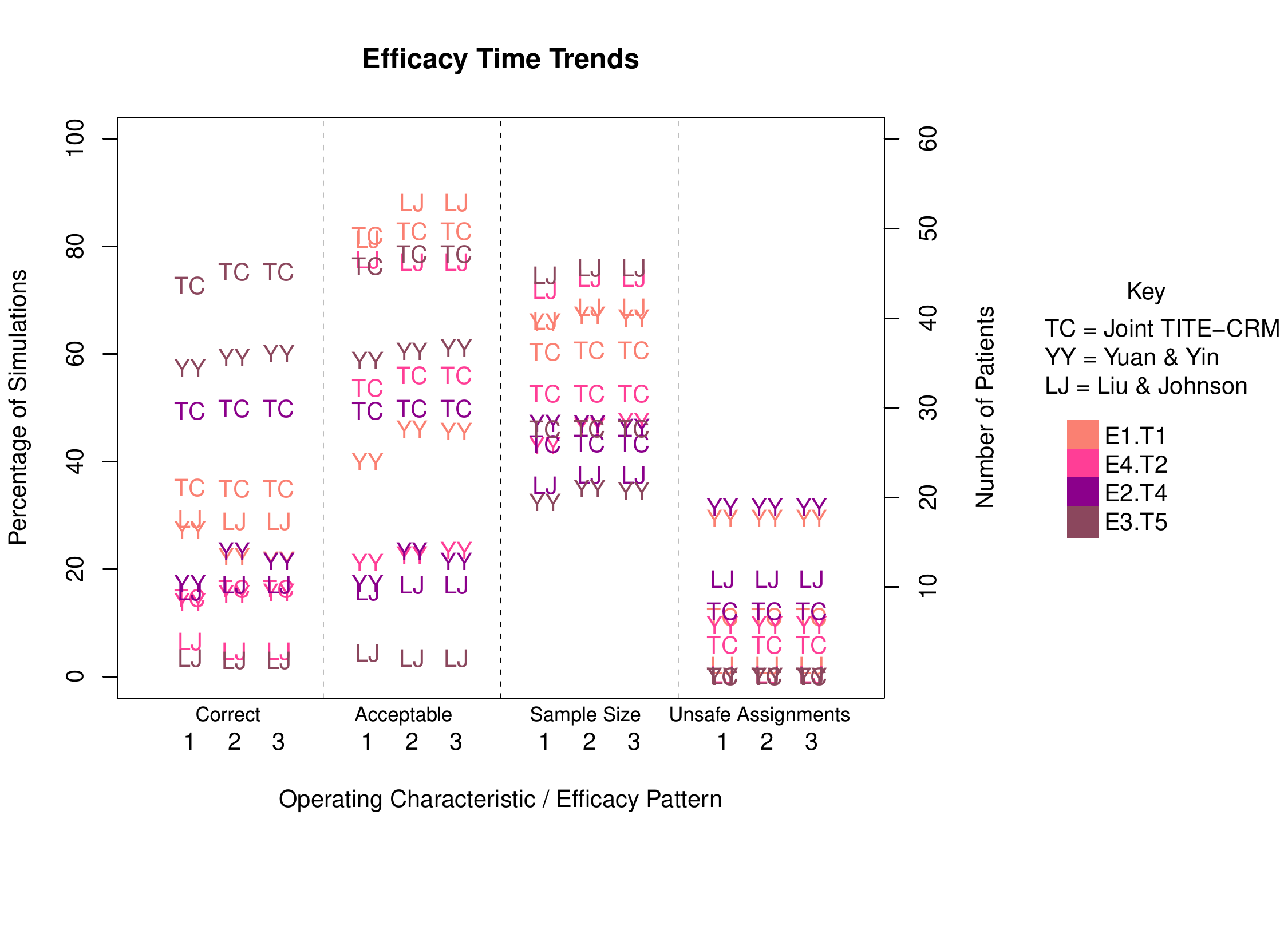}
     \caption{Operating characteristics for the three efficacy time trends}   \label{fig:eff_trends}
  \end{figure}

Figure~\ref{fig:eff_trends} shows the operating characteristics of the three time to event methods when the time trend is varied. There is no difference in the number of patients assigned to unsafe doses, and very little difference to the overall sample size. In terms of selections, the $AT/AE$ methods of Yuan \& Yin is most variable to the time trend, since this method is more heavily dependent on the survival curve. The Joint TITE-CRM and the model-assisted method of Liu \& Johnson are more robust to efficacy time trends.

\section{Discussion} \label{sec:discussion}
In this work, we have compared three designs for a dose-finding trial that uses both efficacy and toxicity outcomes in the form of time-to-event responses. The comparison has highlighted the impact of both the inference of dose-response relationship and the decision criteria. The challenge presented by a trial using both toxicity and efficacy outcomes is compounded by the late onset nature of the responses.\\

The decision criteria used in the $AT/AE$ design, whilst intuitive for survival outcomes, does somewhat deviate from the objective of the dose-finding trial. Without penalty for unsafe doses, it gave an aggressive escalation path in many scenarios. This leads on from the point made regarding Table~\ref{tab:scens_ut}, that without targeting any safety level, scenarios with similar true ratios across levels present very challenging for this design. Additionally, the increased computational intensity of this design is not rewarded by an increase in performance.\\

The model-assisted method and the Joint TITE-CRM implemented both used the same utility criterion, which allowed a comparison between the simple inference of Liu \& Johnson \cite{Liu2016} and the joint logistic model. Here the increase in complexity is rewarded with increased performance. The Joint TITE-CRM is recommended as an alternative that balances complexity and performance across a range of scenarios.

\end{document}